\documentclass[a4paper,11pt]{article}
\usepackage{jinstpub} % for details on the use of the package, please see the JINST-author-manual
\usepackage{orcidlink} 
\title{\boldmath Performance and Radiation Damage of the Heavy-Ion Silicon Beam Tracker in the BM@N Experiment}

\author[1]{D.~Chemezov\,\orcidlink{0009-0003-8562-3126},}
\author{S.~Khabarov\,\orcidlink{0009-0003-8500-1749},}
\author{Yu.~Kopylov\,\orcidlink{0009-0001-4639-8937},}
\author{I.~Osokin\,\orcidlink{0009-0006-3410-2699},}
\author{A.~Sheremeteva\,\orcidlink{0009-0005-5925-3824},}
\author{E.~Streletskaya\,\orcidlink{0009-0003-0627-0175},}
\author{O.~Tarasov\,\orcidlink{0009-0002-0590-524X},}
\author{E.~Zubarev\,\orcidlink{0009-0001-3239-795X}}
\author{and N.~Zamiatin\,\orcidlink{0009-0002-5256-1937}}

\affiliation{Joint Institute for Nuclear Research (JINR), Dubna, Russia}

\note{Corresponding author.}
\emailAdd{chemezov@jinr.ru}

\abstract{The Silicon Beam Tracker (SiBT) of the BM@N experiment at the NICA complex (JINR, Dubna) is designed to reconstruct trajectories of beam ions upstream of the target. It consists of three stations based on double-sided silicon strip detectors (DSSDs) installed directly in the relativistic heavy-ion beam. The SiBT performance and radiation damage were studied using data from the 2026 experimental run with $^{124}$Xe beams at~$1.6$,~$2.2$, and~$3.0$~AGeV. During the run, the leakage current of the DSSD of the third station of SiBT increased by $28.95~\mu\mathrm{A}$, while the charge collection efficiency in the beam center decreased to approximately $68\%$ at the nominal bias voltage of 30~V. Post-run measurements showed that the CCE in the beam center saturates at $\approx93\%$ above $\approx70~\text{V}$. Despite the radiation-induced signal degradation, the SiBT station 2 tracking efficiency remained close to $99\%$. The accumulated radiation damage was also used to reconstruct the spatial distribution of the integrated $1$~MeV neutron-equivalent fluence from strip-by-strip leakage-current measurements, yielding an average value of approximately $1.3\times10^{12}~\mathrm{n_{eq}/cm^2}$.}


\arxivnumber{2610.17947} % Only if you have one

\begin{document}
\maketitle
\flushbottom

\section{Introduction}
\label{sec:intro}
\begin{figure}[htbp]
\centering
\includegraphics[width=.95\textwidth]{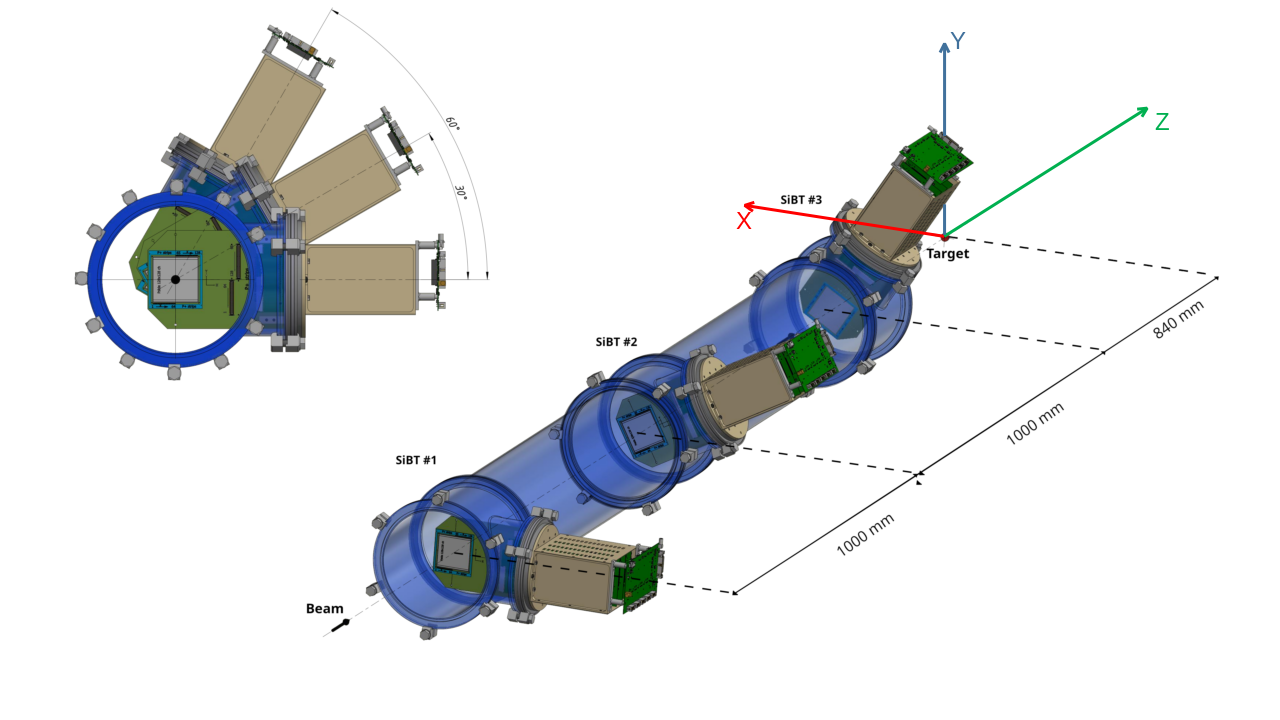}
\caption{Schematic representation of SiBT location in beamline.\label{fig:beam_pipe}}
\end{figure}

The BM@N (Baryonic Matter at Nuclotron) is the first large experiment operating at the Nuclotron-based Ion Collider fAcility (NICA) complex at JINR, Dubna. It is dedicated to studying the interactions of relativistic heavy-ion beams with fixed targets in the energy range of 1 to 6 AGeV \cite{kapishin2019studies}. A general overview of the BM@N experimental setup and the role of the SiBT within it is provided in Ref.~\cite{bmn_spectrometer}. Accurate event kinematics reconstruction and calibration of the main BM@N tracking detectors require trajectory measurement of beam ions and determination of the beam impact angle on the target. 
The Silicon Beam Tracker (SiBT) subsystem is built upon double-sided silicon microstrip detectors (DSSDs) and has three planes separated by about $1$~m along the beam axis. DSSDs are installed directly inside the beam vacuum chamber of the relativistic heavy-ion beamline. Figure~\ref{fig:beam_pipe} shows a schematic of the three SiBT stations mounted in the vacuum beam pipe. 

Two experimental runs with $^{124}$Xe beams were conducted with the SiBT in operation. Run 8 (December 2022 -- February 2023) utilized beam energies of 3.0 and 3.8 AGeV with a CsI target; preliminary results on SiBT performance from this run were reported in Ref.~\cite{proceeding}. Run 9 (January -- April 2026) employed beam energies of 1.6, 2.2, and 3.0 AGeV with a Sn target. This paper presents the SiBT performance during Run 9, including the observed radiation-induced changes.

\section{Silicon Beam Tracker system}
\label{sec:description}
\subsection{Detector Modules and Geometry}
To minimize multiple scattering of beam ions and mitigate radiation-induced noise, the SiBT employs thin double-sided silicon strip detectors. Each detector is fabricated from high-resistivity ($\rho > 5\ \text{k}\Omega\cdot\text{cm}$) 4-inch float-zone (FZ) mono-crystalline silicon wafers. The wafer thickness of $175\ \mu\text{m}$ represents the practical minimum for planar technology on 4-inch wafers. This reduced thickness directly decreases the volume of the space charge region, which is important for maintaining an acceptable noise level per strip.

\begin{figure}[htbp]
\centering
\includegraphics[width=.35\textwidth]{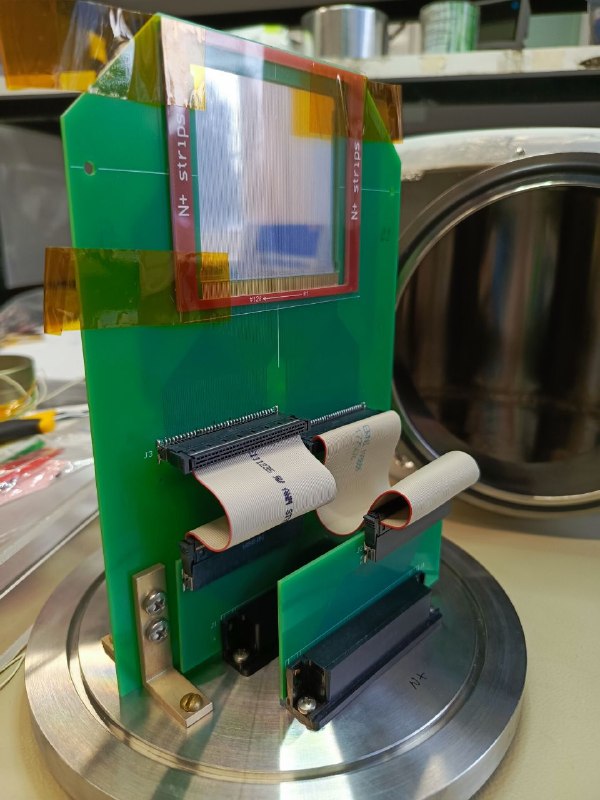}
\caption{Photograph of a SiBT module showing DSSD sensor, mounted on the vacuum flange.
\label{fig:module_photo}}
\end{figure}

Each DSSD features an active area of $61 \times 61\ \text{mm}^2$, segmented into 128 strips on both the p$^+$ (boron-implanted) and n$^+$ (phosphorus-implanted) sides, with a strip pitch of $470\ \mu\text{m}$. The strips on opposite sides are oriented orthogonally. To reduce tracking ambiguities, each subsequent SiBT station is rotated around the beam axis by $30^\circ$ relative to the previous one (Figure~\ref{fig:beam_pipe}). Structurally, the silicon sensors are mounted on printed circuit boards (PCBs) featuring gold contact pads, interconnected to the Al-plated strips on the DSSD via ultrasonic bonding. A photograph of a fully assembled SiBT module is shown in Figure~\ref{fig:module_photo}.

\subsection{Readout Electronics}
\label{sec:readout}
\begin{figure}[htbp]
\centering
\includegraphics[width=.95\textwidth]{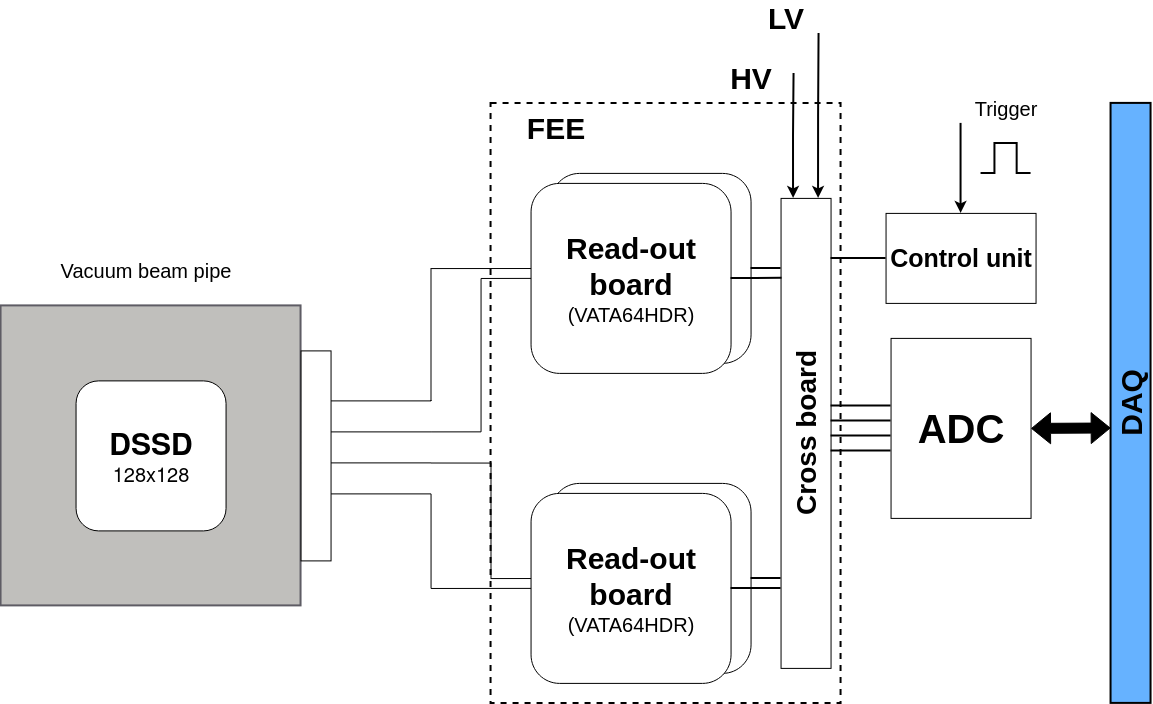}
\caption{Simplified block diagram of the SiBT readout chain.
\label{fig:sibt_readout}}
\end{figure}

A simplified block diagram of the SiBT readout chain is shown in Figure~\ref{fig:sibt_readout}. The signals from the detector strips are grouped into four 64-channel bundles and routed via flat cables to vacuum connectors. Four readout boards, servicing all 256 channels ($64\times2$ p$^+$ and $64\times2$ n$^+$) per station, are mounted on the external side of the vacuum flange~\cite{sibt_siprof_development}. Readout mezzanine cards with wire-bonded and encapsulated ASICs are plugged into the readout boards. All four readout boards are connected to a single cross board, which also distributes low-voltage (LV) power to the readout boards, distributes control and readout sequence from control unit to all ASICs, routes the high-voltage (HV) sensor bias through them to the DSSDs. This topology not only protects the electronics from radiation damage but also allows for in situ testing, tuning, and replacement without breaking the beamline vacuum.

The readout is based on the VATA64HDR16.2 ASIC (IDEAS, Norway)~\cite{ideas}. This ASIC was selected for its wide dynamic range ($-20\ \text{pC}$ to $+55\ \text{pC}$), which is essential for handling the large charge deposits from highly ionizing heavy ions: a Xe ion at the energies used in Run 9 (1.6--3.0 AGeV) traversing $175\ \mu\text{m}$ of silicon generates approximately $11\ \text{pC}$ of charge \cite{bmn_spectrometer}. Since each ASIC handles 64 channels, four chips are utilized per station (two chips per side). During data acquisition, the sampling time is determined by an external beam trigger, provided by a dedicated set of beam scintillation counters (see Sections 3 and 9 in Ref.~\cite{bmn_spectrometer}). SiBT strip signal amplitudes are stored in internal memory capacitors after pulse shaping. In response to the trigger, the control unit generates a readout sequence — distributed to the FEE readout boards via the cross board — that drives an analog multiplexer, sequentially routing the 64 channels to a single ADC for digitization. The same readout architecture is also used for other BM@N tracking systems (based on silicon strip detectors, gas electron multipliers and cathode strip chambers)~\cite{Ivanova2019}.

\subsection{Mechanical Integration and Alignment}
The DSSDs of the three SiBT stations are mounted inside the beam vacuum chamber. The station positions along the beam axis are $Z_{\rm SiBT1} \approx -283$~cm, $Z_{\rm SiBT2} \approx -183$~cm, and $Z_{\rm SiBT3} \approx -84$~cm relative to the target ($Z=0$), corresponding to a $\sim2$~m SiBT1--SiBT3 lever arm with SiBT2 positioned close to its midpoint. The 3D coordinates of each DSSD relative to the geometric axis of the beam pipe were measured using a NORGAU NVM II-5040D video microscope with an accuracy of $\pm 5\ \mu\text{m}$. 
The modular design of the SiBT stations allows the replacement of individual irradiated sensors. Although this procedure requires breaking and re-establishing the beam vacuum, the exchange of a single DSSD module is straightforward and can be performed without dismantling the entire SiBT assembly.
\section{Silicon Beam Tracker performance in Run 9}
\label{sec:performance}
\subsection{Beam Profile and Vertex Reconstruction}
\begin{figure}[htbp]
\centering
\includegraphics[width=.95\textwidth]{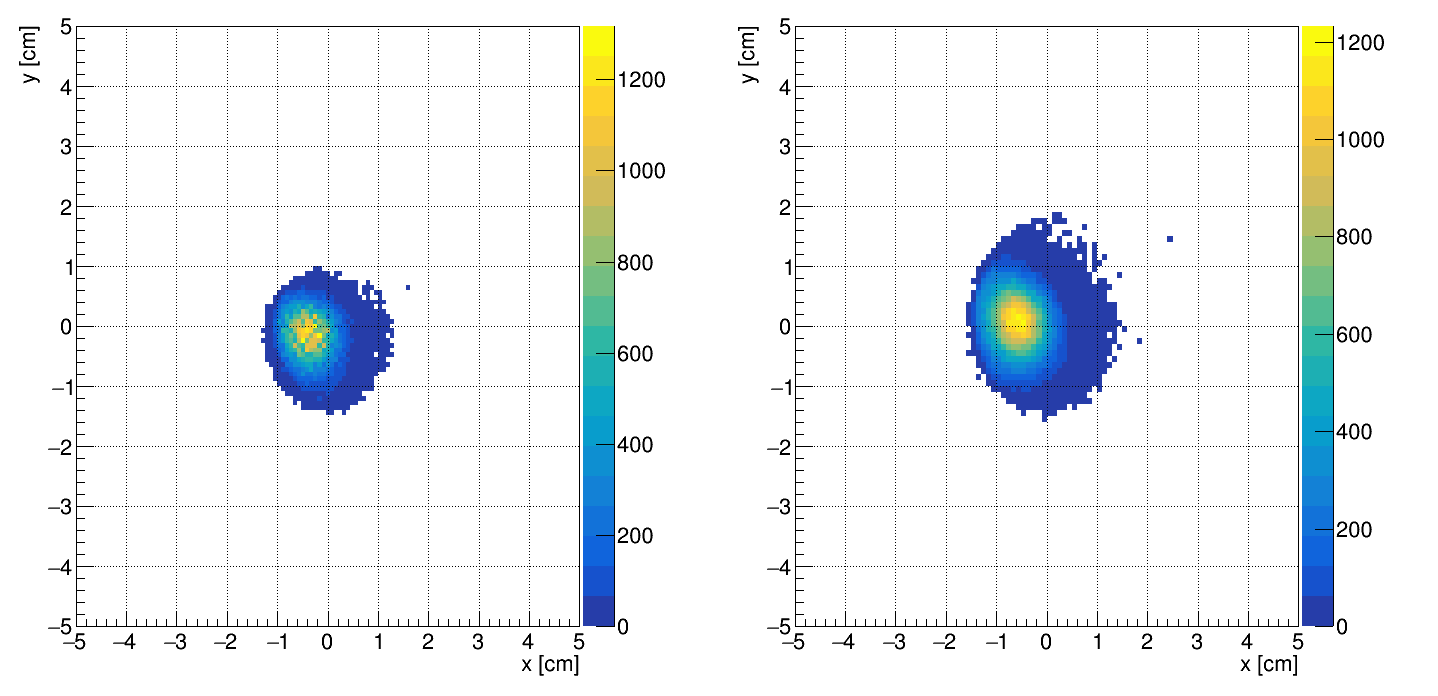}
\caption{Two-dimensional beam profile from SiBT2 (left) and distribution of reconstructed vertex
position at the target plane (right) for $2\times10^6$  events, $E=$~2.2~AGeV.
\label{fig:profile_vertex}}
\end{figure}

The primary function of the SiBT is to provide online monitoring of the heavy-ion beam profile and offline reconstruction of the primary vertex position. Figure~\ref{fig:profile_vertex} shows distributions from $2\times10^6$ events: (left) the two-dimensional beam profile reconstructed from SiBT2 hit positions, and (right) the two-dimensional distribution of beam-particle vertices, extrapolated to the target plane ($z=0$) using the reconstructed SiBT tracks. The SiBT beam-profile distributions were used during the run for beam tuning. The SiBT vertex reconstruction does not determine the $Z$ coordinate of the vertex; it extrapolates tracks to the known target plane. The vertex coordinates obtained from SiBT can be used as an initial approximation for the primary vertex reconstruction algorithm implemented in the internal BM@N tracker.

\subsection{Signal-to-noise ratio}
\label{sec:sig_noise}
The signal-to-noise ratio was evaluated using reconstructed clusters in each station of SiBT. The analysis was performed separately for each readout ASIC (see Section~\ref{sec:readout}), using a sample of $300\,000$ events from the beginning of Run~9 ($2.2$~AGeV period).
The signal-to-noise ratio of each cluster was calculated as:
\begin{equation}
\mathrm{SNR} = \frac{S_{\mathrm{cluster}}}{\sqrt{N_{\mathrm{st,side,1}}^{2}\,n_{1} 
+ N_{\mathrm{st,side,2}}^{2}\,n_{2}}},
\end{equation}
where $S_{\mathrm{cluster}}$ is the sum of the strip signals forming the cluster, $N_{\mathrm{st,side,1}}$ is the mean noise of station $st$, side $side$, ASIC~1 (strips 1--64), $n_{1}$ is the number of strips connected to ASIC~1 that contribute to the cluster, and $N_{\mathrm{st,side,2}}$ and $n_{2}$ are the corresponding quantities for ASIC~2 (strips 65--128). The mean ASIC noise was calculated as a weighted average of the individual channel noises, with the number of cluster hits per channel used as the weight.

\begin{figure}[htbp]
\centering
\includegraphics[width=.95\textwidth]{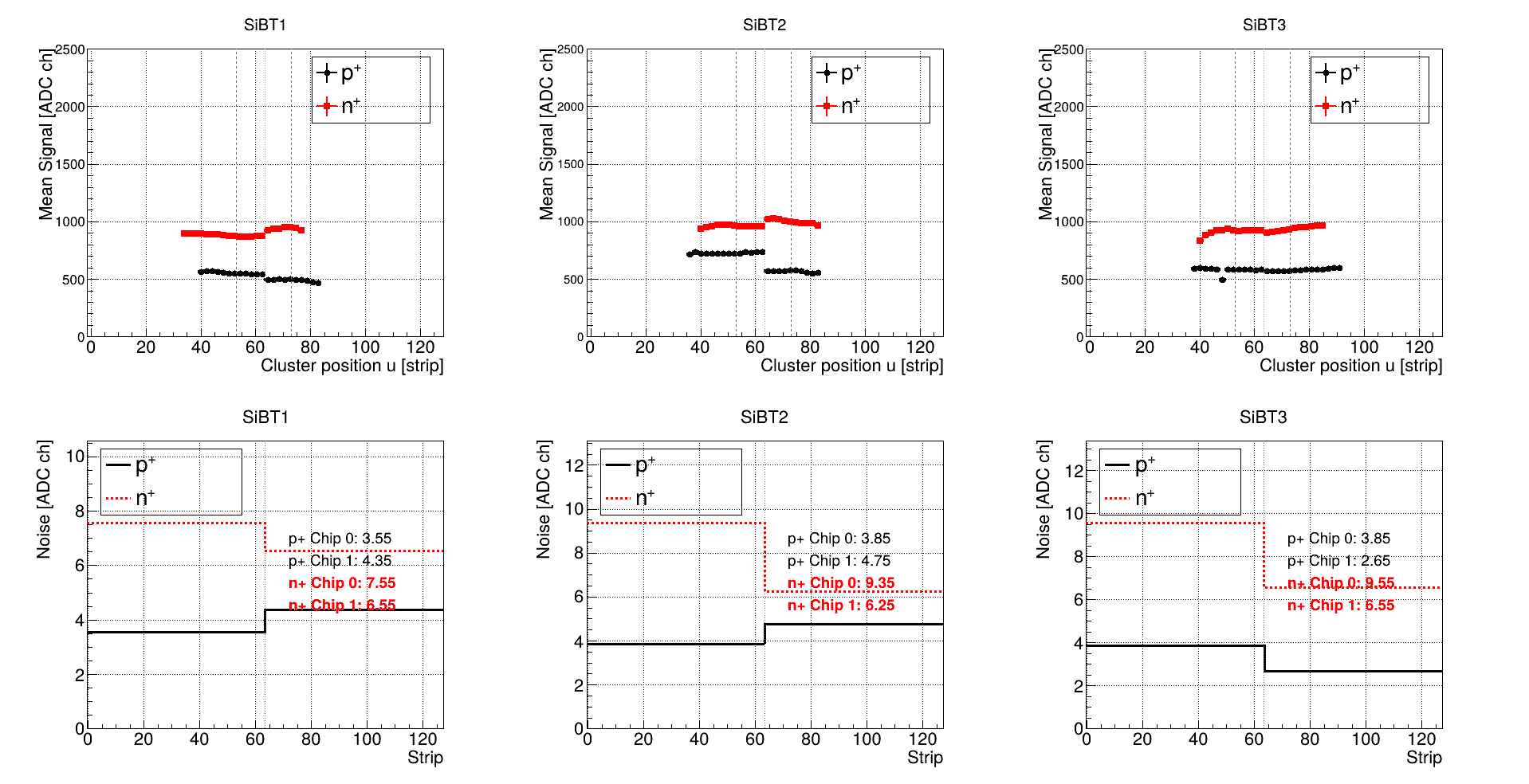}
\caption{Mean cluster signal distribution (top) and mean ASIC noise (bottom) at the beginning of Run~9 for all three SiBT stations. Dashed grey lines indicate the beam spot region used for Table~\ref{tab:sn_begin}.
\label{fig:sn_start}}
\end{figure}

Figure~\ref{fig:sn_start} shows the distributions of mean cluster signal 
(top) and mean ASIC noise (bottom) for all three SiBT stations at the 
beginning of Run~9.

To characterize the detector response in the beam region, clusters with 
positions corresponding to strips 53--73 were selected. 
Table~\ref{tab:sn_begin} presents the weighted mean SNR across the beam 
spot for each ASIC, calculated as an entries-weighted average over the 
position bins within the beam spot.
\begin{table}[htbp]
\centering
\caption{Beam spot SNR at the beginning of Run~9 (strips 53--73).}
\label{tab:sn_begin}
\smallskip
\begin{tabular}{c|c|c|c}
\hline
Station & Side & ASIC & $\langle$SNR$\rangle$ \\
\hline
SiBT1 & p$^+$ & 1 & 154.7\\
SiBT1 & p$^+$ & 2 & 116.0\\
SiBT1 & n$^+$ & 1 & 118.2\\
SiBT1 & n$^+$ & 2 & 144.9\\
\hline
SiBT2 & p$^+$ & 1 & 190.8\\
SiBT2 & p$^+$ & 2 & 122.5\\
SiBT2 & n$^+$ & 1 & 104.1\\
SiBT2 & n$^+$ & 2 & 162.2\\
\hline
SiBT3 & p$^+$ & 1 & 154.0\\
SiBT3 & p$^+$ & 2 & 215.1\\
SiBT3 & n$^+$ & 1 & 98.7\\
SiBT3 & n$^+$ & 2 & 142.3\\
\hline
\end{tabular}
\end{table}

\subsection{Tracking Efficiency and Spatial Resolution}
\label{sec:eff_res}

\textbf{Methodology.}
The tracking efficiency and spatial resolution of SiBT2 were determined using a track interpolation method with a track-unbiased reference trajectory. SiBT1 and SiBT3 were used as the outer reference layers to build a two-point track, while SiBT2, located between them ($Z=-183$~cm), served as the layer under study. Since SiBT2 hit information is not used in the reference track construction, the reference trajectory is independent of the SiBT2 response.

A straight line was fitted through the SiBT1 and SiBT3 hits. The SiBT stations are located in the fringe field of the BM@N analyzing magnet (see Section 2.4 in \cite{bmn_spectrometer}), so the ion trajectories are not exactly straight in this region. The straight-line approximation therefore introduces an additional, unquantified systematic uncertainty that is not corrected in the estimation presented below.

Combinatorial background was suppressed by requiring only one reconstructed hit on each of SiBT1 and SiBT3, since with two space points a fake track cannot otherwise be rejected.
The efficiency and resolution were then extracted with an iterative, two-pass procedure:
\begin{enumerate}
\item \emph{First pass.} The two-point track was interpolated to the SiBT2 plane and matched to its nearest hit. The resulting $X$ and $Y$ residual distributions were fitted with a Gaussian to obtain initial resolution estimates, $\sigma_x^{\rm init}$ and $\sigma_y^{\rm init}$; the track was also extrapolated to the target plane ($z=0$), and the resulting $X_{\rm vert}$ and $Y_{\rm vert}$ distributions were fitted with a Gaussian to obtain the most probable vertex position and its $\sigma$.
\item \emph{Second pass.} In the second pass, only tracks whose vertex coordinate fell within $\pm2\sigma$ of the most probable vertex position were retained. A SiBT2 hit was counted as a successful match if the distance to the track, normalized to the first-pass resolution,
\begin{equation}
dR = \sqrt{\left(\frac{X_{\rm hit}-X_{\rm track}}{\sigma_x^{\rm init}}\right)^{2} + \left(\frac{Y_{\rm hit}-Y_{\rm track}}{\sigma_y^{\rm init}}\right)^{2}}
\label{eq:dR}
\end{equation}
satisfied $dR \le 5$. The residuals of the matched hits were re-fitted with a Gaussian to obtain the final spatial resolution.
\end{enumerate}
The tracking efficiency was defined as the fraction of selected tracks with a matched ($dR\le5$) hit on SiBT2, with statistical uncertainty
\begin{equation}
\delta_{\varepsilon} = \sqrt{\varepsilon(1-\varepsilon)/N_{\rm tracks}},
\label{eq:eff_error}
\end{equation}
where $\varepsilon$ is the efficiency and $N_{\rm tracks}$ is the number of selected tracks.

\begin{figure}[htbp]
\centering
\includegraphics[width=.95\textwidth]{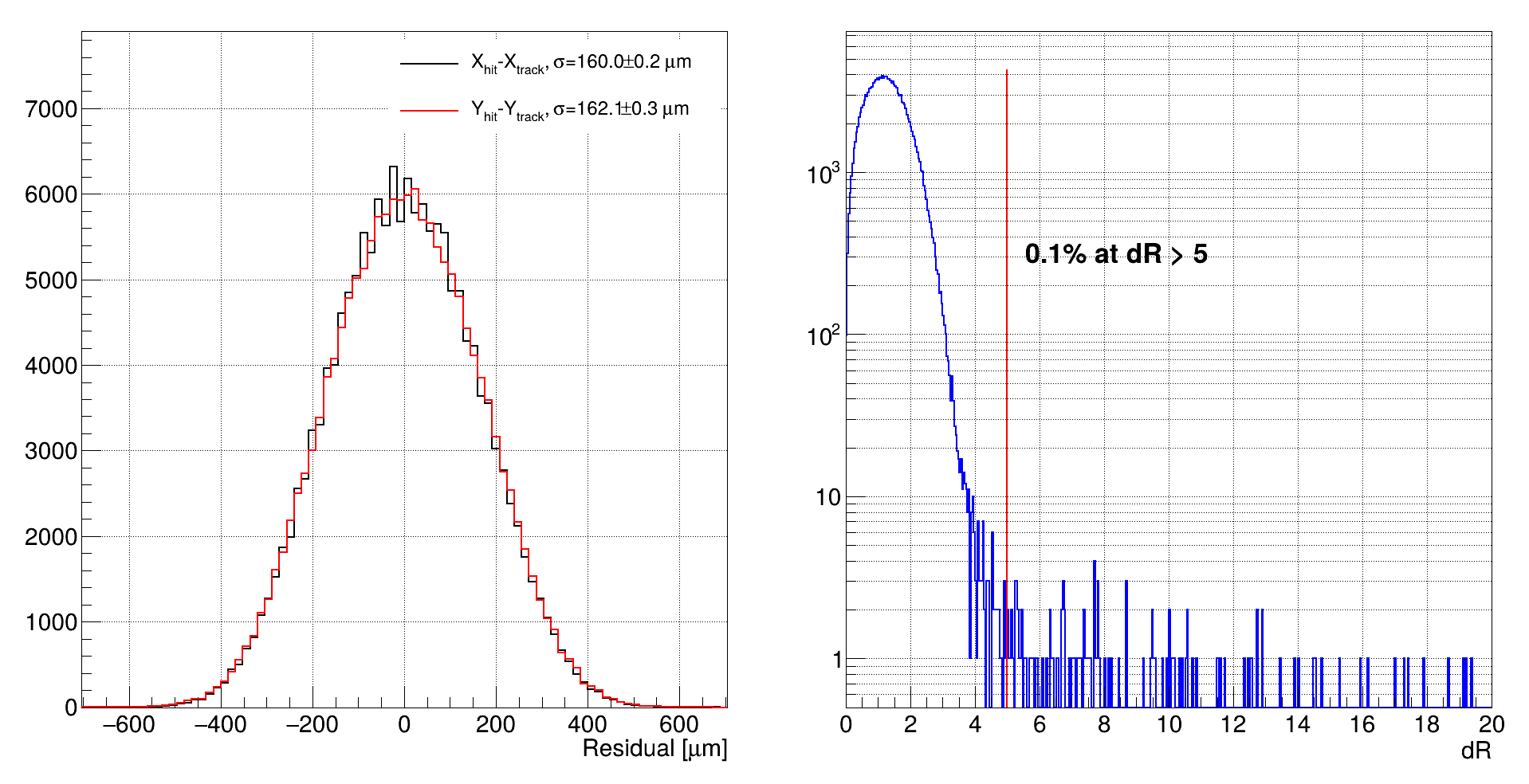}
\caption{SiBT2 $X$ and $Y$ residual distributions of the matched hits (left), and the distribution of the normalized distance $dR$ (right, Eq.~\ref{eq:dR}), for a representative sample of $300\,000$ events at the start of the run ($2.2$~AGeV).\label{fig:eff_res_single_run}}
\end{figure}

\textbf{Results.}
Figure~\ref{fig:eff_res_single_run} illustrates the method on a representative sample of $300\,000$ events from the start of the run ($2.2$~AGeV period): the left panel shows the final $X$ and $Y$ residual distributions of the matched hits, and the right panel shows the distribution of the normalized distance $dR$ (Eq.~\ref{eq:dR}), with the $dR\le5$ selection indicated.

Out of $300\,000$ recorded events, $178\,113$ had exactly one reconstructed hit on each of SiBT1 and SiBT3, rejecting $121\,887$ events ($40.6\%$) with zero or multiple hits. Of these, $159\,796$ had a vertex coordinate within $\pm2\sigma$ of the most probable vertex position and were retained as selected tracks, rejecting a further $18\,317$ events ($10.3\%$ of those passing the hit-multiplicity cut). Of the selected tracks, $159\,229$ satisfied $dR\le5$ ($567$ rejected, $0.35\%$), corresponding to a tracking efficiency of {\boldmath $(99.65\pm0.01)\%$}; the fitted spatial resolution is {\boldmath $\sigma_x=(160.0\pm0.2)\ \mu\textbf{m}$} and {\boldmath $\sigma_y=(162.1\pm0.3)\ \mu\textbf{m}$}. 
The evolution of the tracking efficiency, spatial resolution and signal-to-noise ratio over the full run, and their sensitivity to the radiation damage accumulated by the sensors, are discussed in Section~\ref{sec:tracking_stability}.

\section{Radiation damage}
\label{sec:radiation}
The SiBT DSSDs, located in the direct beam of relativistic heavy ions, were continuously irradiated by Xe ions. In this section, we present an analysis of radiation-induced damage using the SiBT3 detector as a representative example. The radiation effects in SiBT1 and SiBT2 are expected to be similar, since the three stations are identical and were operated in the same beam under the same conditions; a comparable SNR degradation was observed in all three stations (Section~\ref{sec:tracking_stability}).
\subsection{Leakage Current Evolution}

\begin{figure}[htbp]
\centering
\includegraphics[width=.95\textwidth]{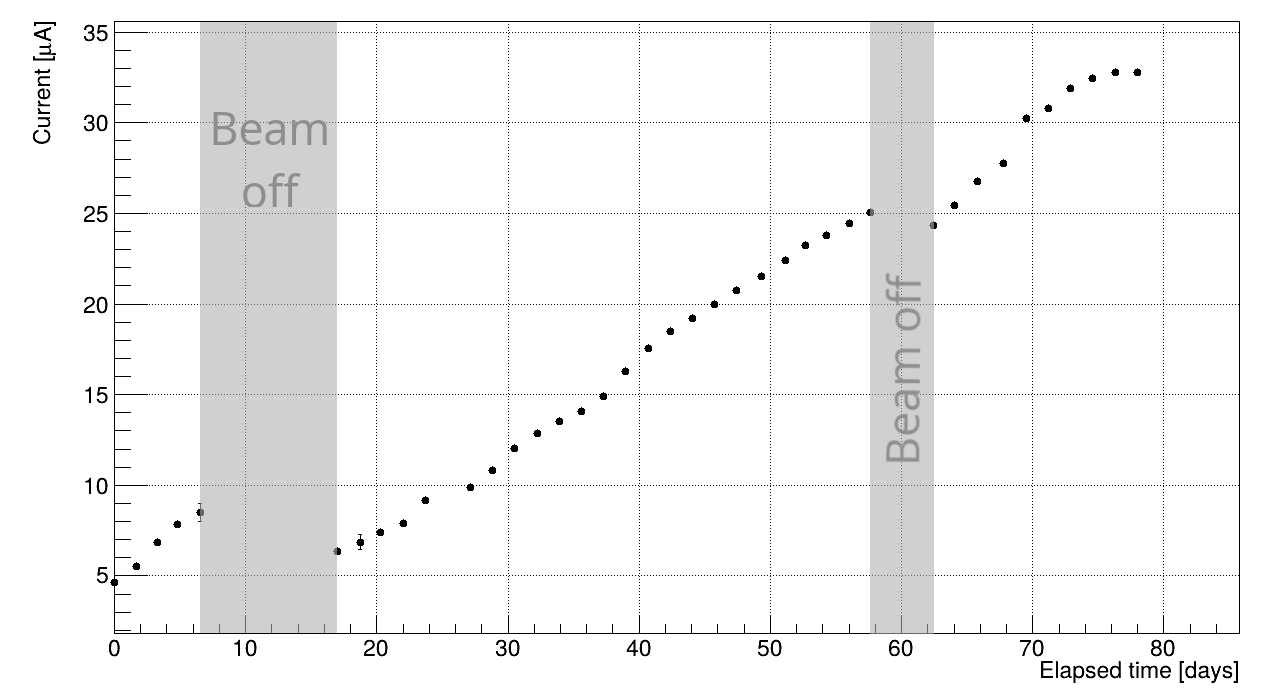}
\caption{Evolution of the leakage currents of the SiBT3 DSSD during the run.\label{fig:all_currents}}
\end{figure}

The most immediate indicator of radiation damage in silicon detectors is the increase in bulk leakage current. Figure~\ref{fig:all_currents} shows the evolution of the SiBT3 leakage current during the run. Over the course of the run, the total leakage current of the SiBT3 detector increased by $\Delta I = 28.95~\mu\text{A}$ (normalized to $20^\circ\text{C}$). For comparison, the current increase during Run 8 was only $6.7~\mu\text{A}$, reflecting the higher integrated beam intensity and exposure time in the recent campaign.

\begin{figure}[htbp]
\centering
\includegraphics[width=.95\textwidth]{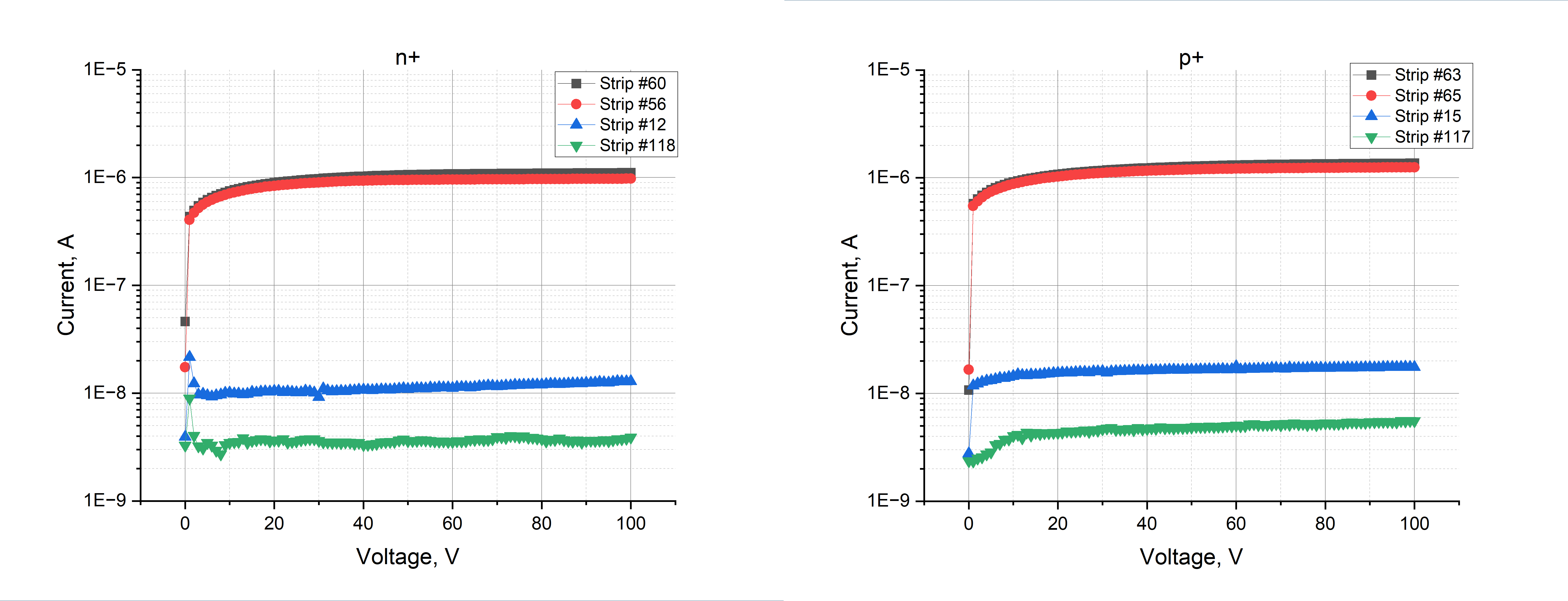}
\caption{SiBT3 individual I-V characteristics for $n^+$ (left) and $p^+$ (right) strips.\label{fig:iv_curves}}
\end{figure}

After the run, I-V characteristics were measured for individual strips on both p$^+$ and n$^+$ sides. Figure~\ref{fig:iv_curves} shows the comparison between heavily irradiated central strips (strip 60 and strip 56 for n$^+$ side; strip 63 and strip 65 for p$^+$ side) and weakly irradiated edge strips (strip 12 and strip 118 for n$^+$ side; strip 15 and strip 117 for p$^+$ side). The irradiated strips exhibit a clear increase in leakage current.

The values of the detector leakage currents and the most damaged strips on the n$^+$ and p$^+$ sides at a voltage of 30 V are shown in Table~\ref{tab:leakage_currents}.
\begin{table}[htbp]
\centering
\caption{Leakage currents (at 20$^\circ$C) before and after Run 9.\label{tab:leakage_currents}}
\smallskip
\begin{tabular}{c|c|c}
\hline
 &$I_{start}\ \mu \text{A}$ & $I_{end}\ \mu \text{A}$\\
\hline
SiBT3 & 3.85 & 32.8\\
p$^+$ strip 63 & 0.153 & 1.140\\
n$^+$ strip 60 & 0.138 & 1.331\\
\hline
\end{tabular}
\end{table}
\subsection{Signal Amplitude Degradation}
\begin{figure}[htbp]
\centering
\includegraphics[width=.65\textwidth]{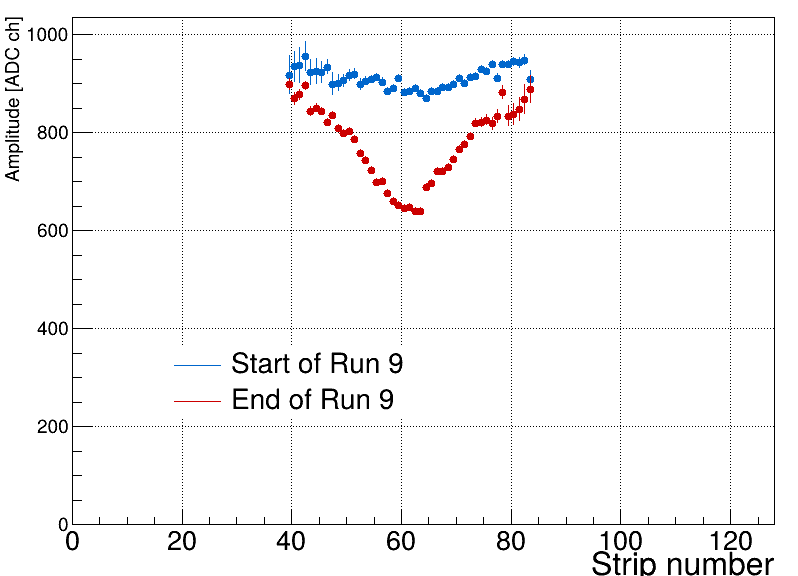}
\caption{Evolution of mean signal amplitude of SiBT3 n$^+$ strips.\label{fig:amplitude_comparison}}
\end{figure}

\begin{figure}[htbp]
\centering
\includegraphics[width=.65\textwidth]{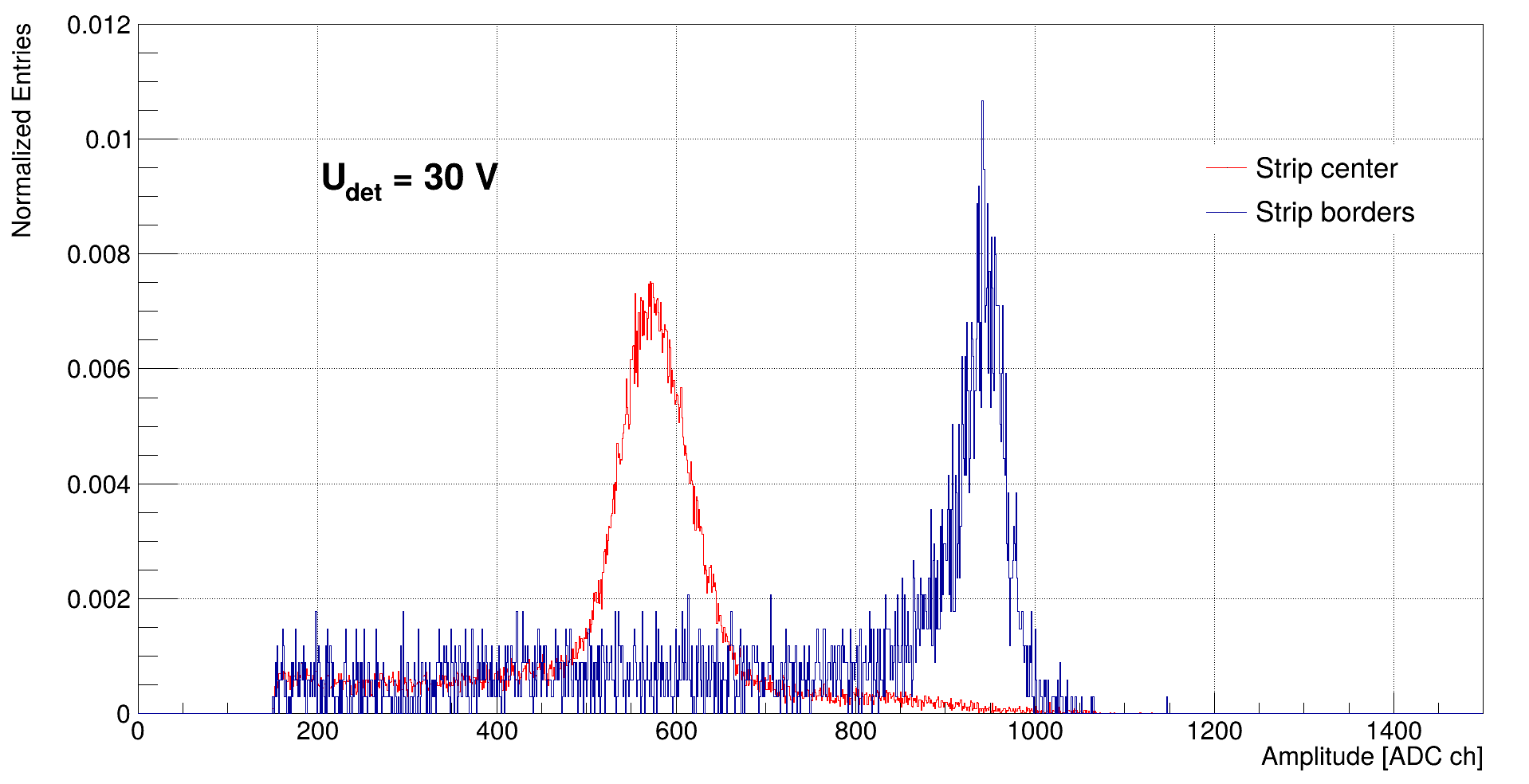}
\caption{Comparison of the amplitudes of the SiBT3 n$^+$ strip 60 in the central and peripheral areas of the beam spot at the end of the run (4 April 2026, $3.0$~AGeV).\label{fig:strip_segmentation}}
\end{figure}

During the run, the signal amplitude decreased in the central, high-intensity region of the beam. Figure~\ref{fig:amplitude_comparison} shows the mean pulse height distribution across n$^+$ strips at the beginning and end of the experimental run.
To study the spatial dependence within a single strip, we compared the signal amplitudes in different segments of a central n$^+$ strip. The strip was virtually segmented according to its intersections with the p$^+$ strips. Events were selected by requiring hits either in the central region (p$^+$ strips 58--66) or in the peripheral regions (p$^+$ strips 38--42 and 76--80) of the beam spot. Figure~\ref{fig:strip_segmentation} shows that the mean signal amplitude in the central segment ($\sim575~\text{ADC ch}$) is lower by a factor of approximately 1.6 than in the peripheral segments ($\sim943~\text{ADC ch}$).
\subsection{Charge Collection Efficiency Measurements}
After the run, the SiBT3 DSSD was removed from the beam pipe and characterized with an $^{241}\text{Am}$ alpha source incident on the n$^+$ side. The alpha particles stop within the first $\approx25~\mu\text{m}$ of silicon, so the signal is dominated by holes drifting towards the p$^+$ side across almost the full sensor thickness.

\begin{figure}[htbp]
\centering
\includegraphics[width=.95\textwidth]{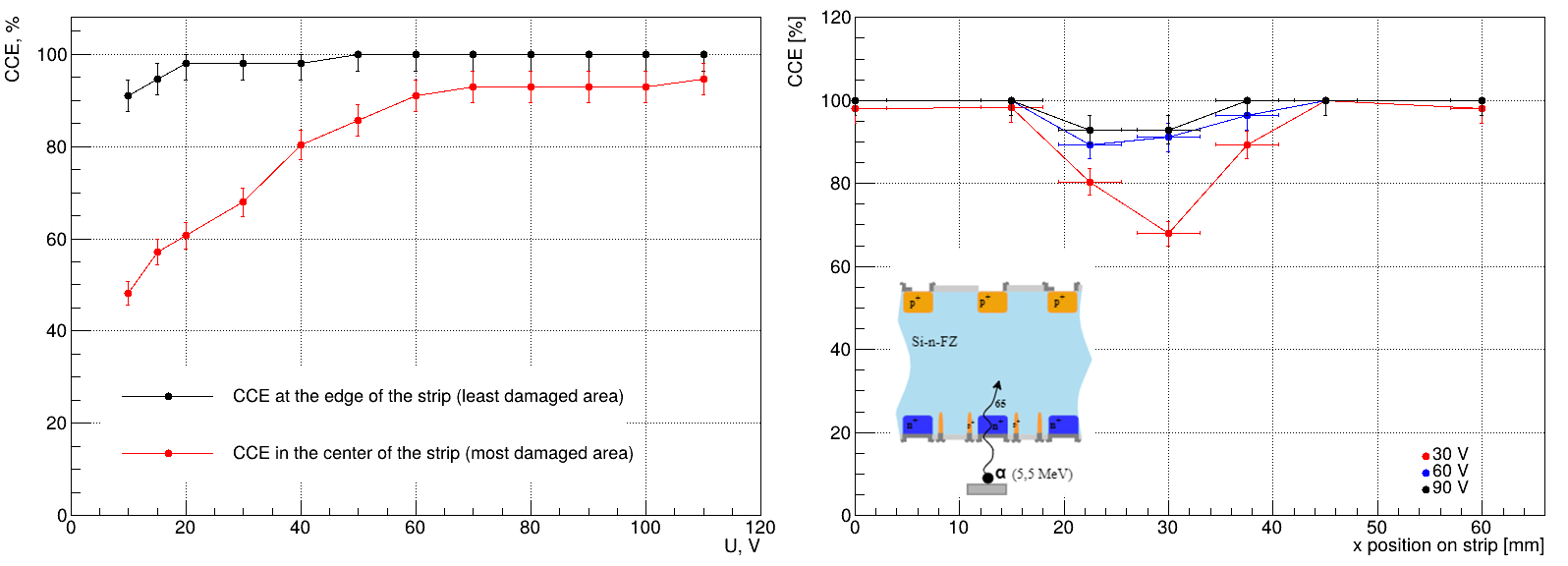}
\caption{Voltage dependence of CCE for the central and edge regions of the strip n$^+$ 65 SiBT3 (left). Dependence of the CCE on the position of the source on the strip n$^+$ 65 SiBT3 (right). The alpha source is incident on the n$^+$ side, as shown schematically.\label{fig:cce_measurements}}
\end{figure}

\textbf{CCE vs. Bias Voltage.} Figure~\ref{fig:cce_measurements} (left) shows the CCE versus bias voltage for the edge (weakly irradiated) and central (heavily irradiated) regions of strip n$^+$ 65, normalized to the edge plateau. The edge saturates already at $\approx20~\text{V}$, whereas the central region approaches its plateau only above $\approx70~\text{V}$, and the plateau itself is reduced to $\approx93\%$. The shifted onset is consistent with an increased effective depletion voltage. The residual $\approx7\%$ loss on the plateau is not removed by further increasing the bias and is consistent with trapping on radiation-induced deep levels. Unambiguous separation of the two contributions would require C--V measurements, which were not performed here.

\textbf{Spatial CCE Profile.} The CCE was also mapped along the strip at $30$, $60$ and $90~\text{V}$ (Figure~\ref{fig:cce_measurements}, right). At the nominal $30~\text{V}$ the loss is confined to the beam core, where the CCE drops to $\approx68\%$ while the edges remain near $100\%$; at $60~\text{V}$ the central value rises to $\approx91\%$. Since a higher bias recovered the CCE in the post-run measurements, the same strategy will be tested under beam conditions in future runs.

\subsection{Spatial Fluence Reconstruction from Leakage Current}
\label{sec:fluence}
\begin{figure}[htbp]
\centering
\includegraphics[width=.75\textwidth]{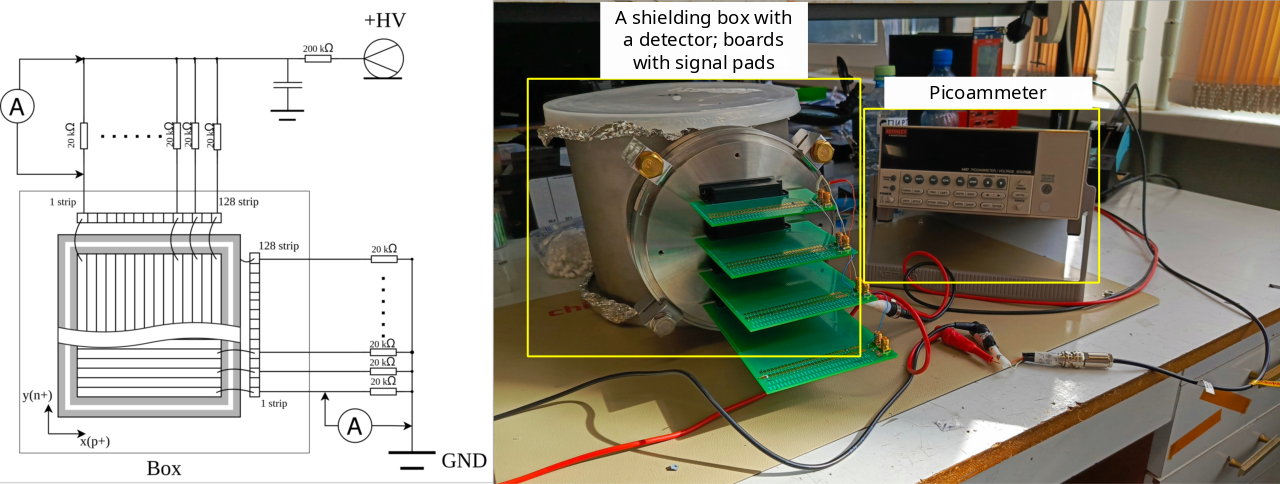}
\caption{Schematic diagram of the leakage current measurement setup (left). Photograph of the measurement setup (right).\label{fig:currents_measurements_setup}}
\end{figure}

\begin{figure}[htbp]
\centering
\includegraphics[width=.95\textwidth]{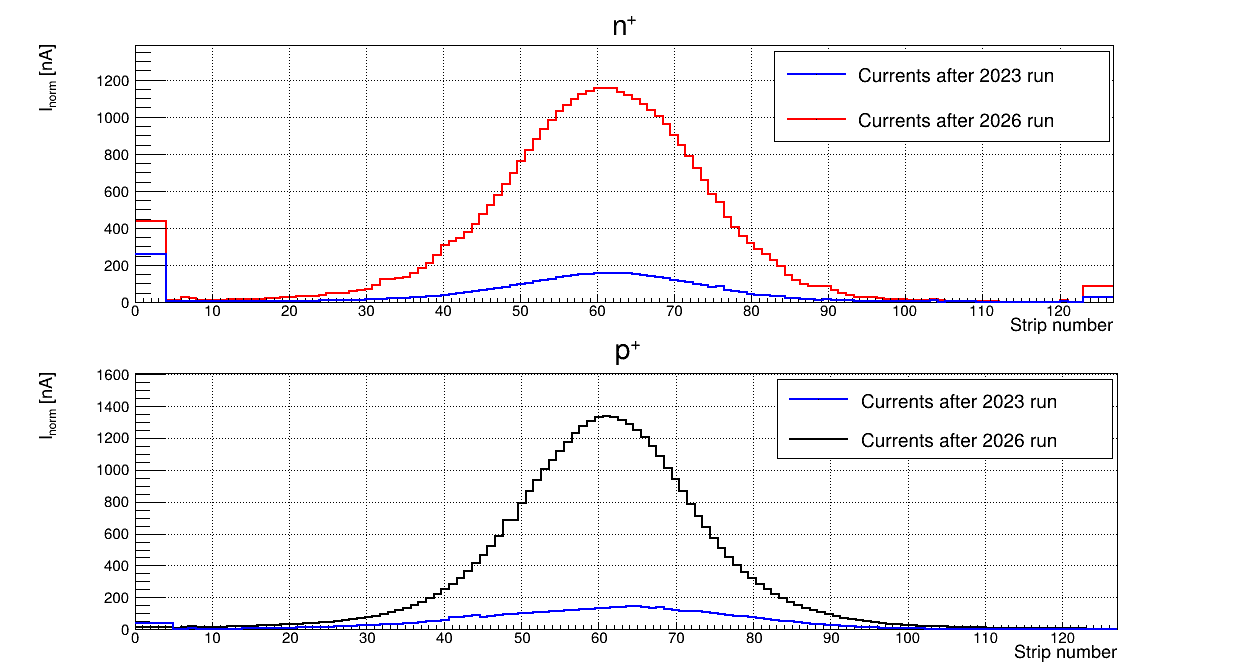}
\caption{Leakage currents of individual SiBT3 strips for n$^+$ (top) and p$^+$ (bottom) sides of the detector.\label{fig:strip_currents}}
\end{figure}

After the experimental run, the leakage currents of individual SiBT3 detector strips were measured. A schematic diagram and a photograph of the current measurement setup are shown in Figure~\ref{fig:currents_measurements_setup}. The measured leakage current distributions are presented in Figure~\ref{fig:strip_currents}. For comparison, similar distributions obtained after Run 8 are also shown.

A pronounced ``radiation memory'' effect can be observed: the leakage current distribution along the detector strips reflects the integral beam profile accumulated during the entire run. Radiation-induced leakage current provides a measure of the accumulated displacement damage. We therefore use the measured strip currents to reconstruct its spatial distribution across the detector. The distribution of leakage currents of strips after Run 8 allows us to find $\Delta I$ for each strip after irradiation during Run 9.

The increase in leakage current is related to the equivalent fluence of 1~MeV neutrons by the empirical relationship \cite{shafronovskaia2014method}:

\begin{equation}
\Delta I = \alpha \cdot \Phi_{eq} \cdot V_{det},
\end{equation}

where $\Delta I$ is the radiation-induced leakage current increase (in our case, it is the difference between the post-Run 9 and post-Run 8 measurements), $\alpha$ is the current-related damage factor for silicon, $\Phi_{eq}$ is the equivalent fluence of 1~MeV neutrons, and $V_{det}$ is the detector volume.

\textbf{Methodology.}
The reconstruction of the spatial distribution of the accumulated equivalent neutron fluence is based on the following procedure:

\begin{enumerate}
\item Each n$^+$ strip is divided into 128 virtual pixels of size $470 \times 470~\mu\mathrm{m}^2$, corresponding to the intersections with p$^+$ strips.

\item A weighting coefficient is assigned to each p$^+$ strip according to its relative contribution to the total detector leakage current:

\begin{equation}
    \omega_i = \frac{I_i}{I_{\Sigma}}, \qquad i = 1 \ldots 128,
\end{equation}

where $I_i$ is the leakage current of the $i$-th p$^+$ strip and $I_{\Sigma}$ is the total detector leakage current.

\item The leakage current increase $\Delta I_j$ of each n$^+$ strip is distributed among its 128 virtual pixels according to the weighting coefficients $\omega_i$.

\item The local accumulated equivalent neutron fluence in each virtual pixel $(i,j)$ is calculated as

\begin{equation}
    \Phi_{eq}(i,j) =
    \frac{\Delta I_j \cdot \omega_i}
    {\alpha \cdot V_{pix}},
\end{equation}

where $\Delta I_j$ is the leakage current increase of the $j$-th n$^+$ strip, $\omega_i$ is the weighting coefficient determined from the p$^+$ strip current distribution, and

\[
V_{pix} = 0.047 \times 0.047 \times 0.0175~\mathrm{cm}^3
\]

is the volume of a virtual pixel.
\end{enumerate}
The leakage currents of individual strips were measured 9 days after the end of the run. The coefficient $\alpha = 3.6 \times 10^{-17}$~A/cm~\cite{moll_phd} was used to convert the current increase into equivalent fluence; the choice of effective temperature and the associated systematic uncertainties are discussed below.

\begin{figure}[htbp]
\centering
\includegraphics[width=.95\textwidth]{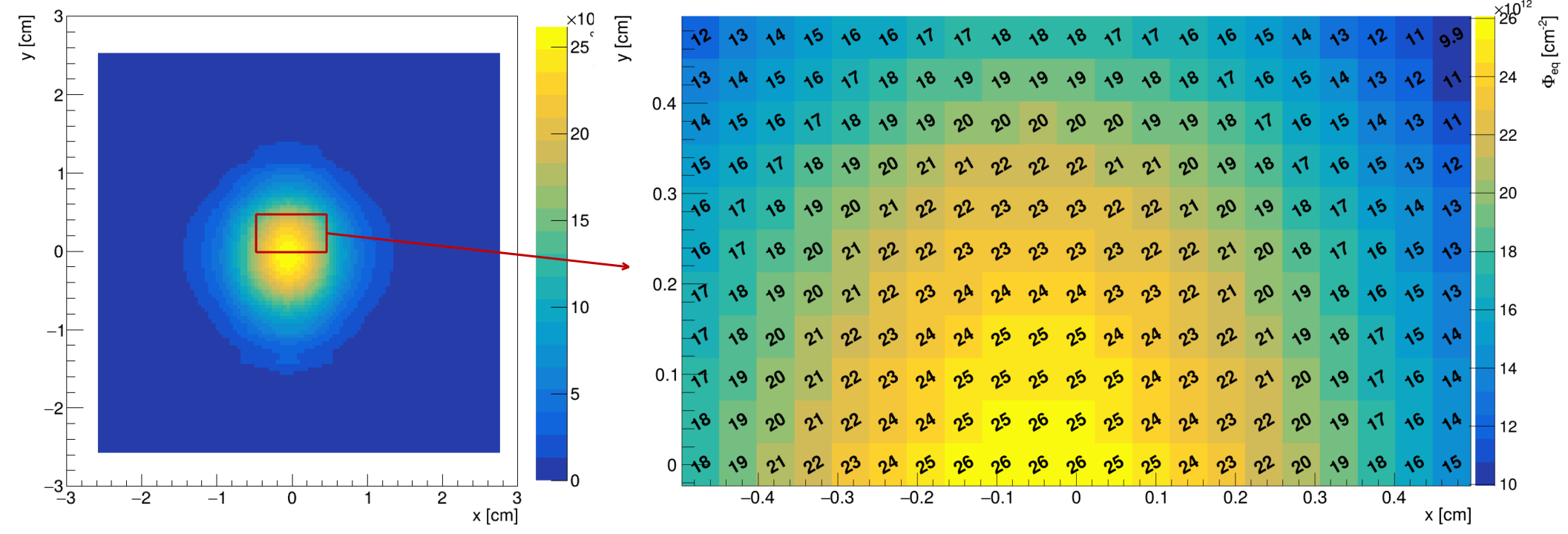}
\caption{Reconstructed spatial distribution of the accumulated 1~MeV neutron equivalent fluence in the SiBT3 detector. On the left is the distribution over the entire DSSD, on the right is the enlarged sector.}
\label{fig:2d_fluence}
\end{figure}

\textbf{Results.}
Figure~\ref{fig:2d_fluence} shows the reconstructed two-dimensional map of the accumulated 1~MeV neutron equivalent fluence.

The average equivalent fluence over the detector area, calculated from the total leakage current increase of $28.95~\mu\mathrm{A}$, was found to be $\Phi_{eq} = 1.23 \times 10^{12}~\mathrm{n_{eq}/cm^2}$.
The average fluence obtained from the reconstructed two-dimensional map was calculated as: 
\[
\Phi_{eq} = \frac{\sum_{i,j}\Phi_{eq}(i,j)\cdot S_{pix}}{S_{det}} = 1.32\times 10^{12}~\mathrm{n_{eq}/cm^2} .
\]
The difference of less than $10\%$ is consistent with the difference between the independently measured total detector current and the sum of the individual strip currents.

\textbf{Uncertainties.}
The reconstructed equivalent fluence is subject to the following systematic uncertainties.
\begin{itemize}

\item \textbf{Annealing during the run.} Irradiation and annealing proceeded simultaneously throughout the 79-day run: damage accumulated early in the run therefore had a longer time to anneal than damage produced near its end. The effective damage factor for this irradiation history is lower than the value obtained for a single 9-day annealing period, resulting in an underestimation of the reported fluence. No correction for this effect is applied.

\item \textbf{Post-run annealing temperature.} The per-strip currents were measured 9 days after the end of the run. The hall temperature was recorded continuously over this period and ranged from $26.2^\circ$C to $34.9^\circ$C, with a time-weighted mean of $27.5^\circ$C. Because annealing is thermally activated, the effective temperature is slightly higher than the time-weighted mean; an Arrhenius-weighted estimate gives $27.7^\circ$C. We therefore adopt $28^\circ$C, corresponding to $\alpha = 3.6\times10^{-17}$~A/cm. The recorded temperature refers to the experimental hall air rather than to the sensor itself, and is used as a first approximation to the sensor temperature.

\item \textbf{Temperature normalization of the currents.} All currents were normalized to $20^\circ$C. The logarithmic sensitivity of the silicon leakage current to temperature is $\sim8\%$ per kelvin in this range. The thermometer (Termex LTA-NTS) accuracy of $0.05$~K contributes negligibly. The dominant uncertainty arises from a possible temperature difference between the sensor and the hall air: assuming a difference of $0.4$~K gives $\sim3\%$ in $\Delta I$.

\item \textbf{Other contributions.} The Run~8 baseline currents were measured more than 18 months after the end of that run, when self-annealing had saturated, using the same setup and procedure as the Run~9 measurements. The remaining uncertainty is dominated by the instrument precision. The per-strip measurements were performed with the detector and the guard ring grounded, which suppresses the surface contribution to the measured current.

\end{itemize}

\textbf{Discussion on NIEL Scaling.}
The fluence values reported in this work are expressed in units of 1~MeV neutron-equivalent fluence. The irradiation was performed with $^{124}$Xe ions at energies of $1.6$, $2.2$, and $3.0$~AGeV. Conversion to the corresponding Xe ion fluence would require the appropriate NIEL hardness factor for this irradiation scenario. According to the documentation of the SR-NIEL calculator~\cite{srniel}, when the hadronic contribution is included, the maximum energy is limited to $1$~GeV/nucleon (and to $24$~GeV for protons in silicon) owing to the availability of hadronic interaction data; the $^{124}$Xe energies used in Run~9 ($1.6$--$3.0$~AGeV) lie above this limit. We therefore do not report a hardness factor, and the present study does not determine it independently. The fluence values given here should not be interpreted as the number of Xe ions per unit area.

It is important to distinguish the spatial distribution from its absolute normalization. Since $\alpha$ and $V_{pix}$ are common to all pixels, they cancel in the normalized map: its shape is fixed by the measured strip currents alone, while $\alpha$ and the NIEL scaling assumption affect only the absolute scale. The reconstructed map is a direct image of the spatial distribution of the integrated ion fluence. The residual systematic in the shape originates from the reconstruction procedure itself, which assumes that the current of each n$^+$ strip can be distributed among its virtual pixels according to the p$^+$ strip weights.

A direct comparison with the number of ions delivered to the detector would considerably strengthen the absolute normalization. No such measurement is available for Run~9, and the implementation of an ion counter near the SiBT is planned for future runs.
\subsection{Impact of Radiation Damage on Tracking Performance}
\label{sec:tracking_stability}
\begin{figure}[htbp]
\centering
\includegraphics[width=.95\textwidth]{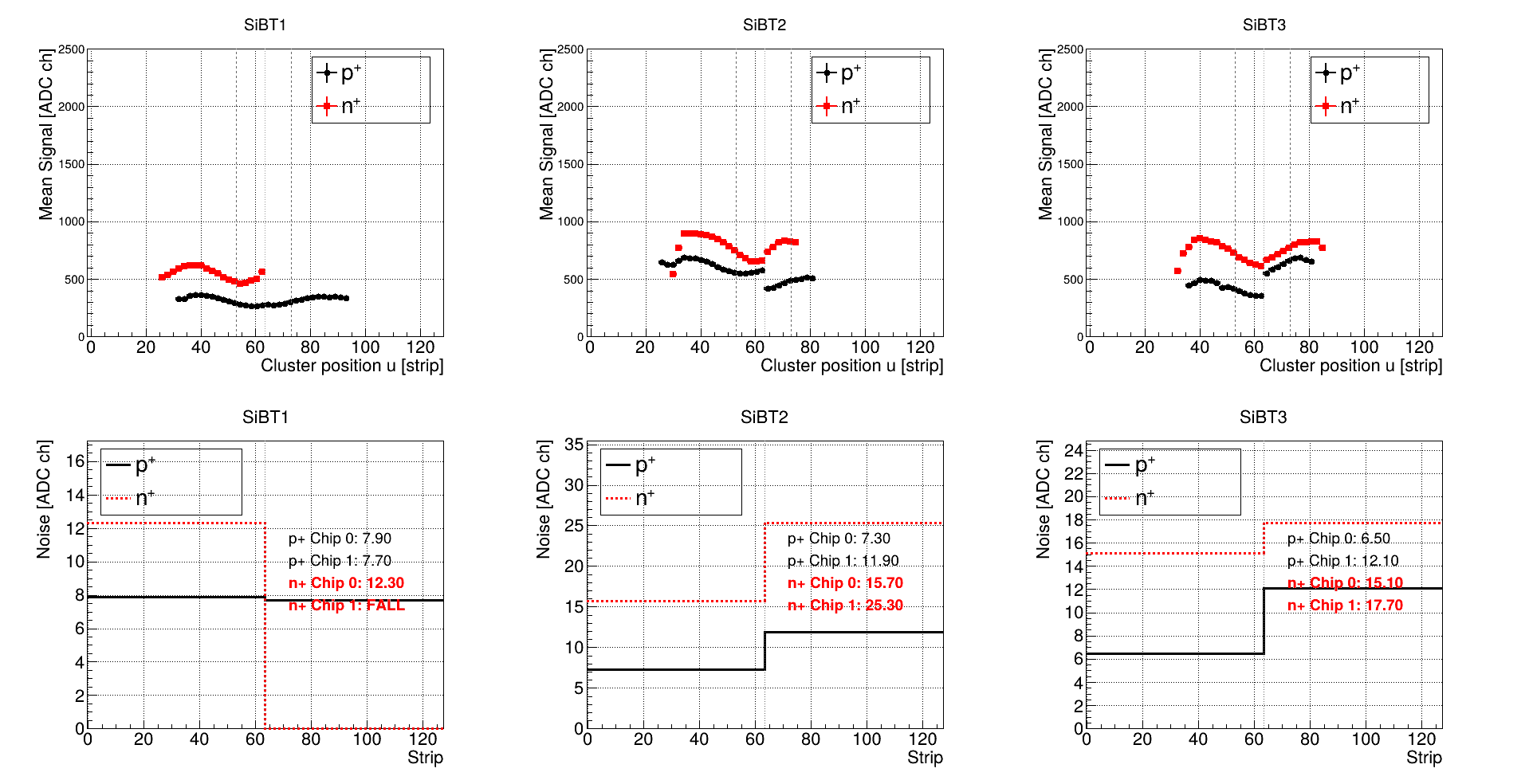}
\caption{Mean cluster signal distribution (top) and mean ASIC noise (bottom) at the end of Run~9 for all three SiBT stations. Dashed grey lines indicate the beam spot region used for Table~\ref{tab:sn_end}.
\label{fig:sn_end}}
\end{figure}

The evolution of the SNR was evaluated using the same methodology as in 
Section~\ref{sec:sig_noise}. Figure~\ref{fig:sn_end} shows the mean cluster signal and mean ASIC noise distributions for all SiBT stations at the end of Run~9, while Table~\ref{tab:sn_end} summarizes the corresponding weighted mean SNR values in the beam spot. One ASIC (SiBT1, n$^+$ side, ASIC~2) failed during the final days of beam delivery and is excluded from the analysis. This ASIC covers strips 65--128, while the beam center was located near strip~53 during the corresponding period.
\begin{table}[htbp]
\centering
\caption{Beam spot SNR at the end of Run~9 (strips 53--73).}
\label{tab:sn_end}
\smallskip
\begin{tabular}{c|c|c|c}
\hline
Station & Side & ASIC & $\langle$SNR$\rangle$\\
\hline
SiBT1 & p$^+$ & 1 & 34.9\\
SiBT1 & p$^+$ & 2 & 36.7\\
SiBT1 & n$^+$ & 1 & 39.0\\
SiBT1 & n$^+$ & 2 & ---\\
\hline
SiBT2 & p$^+$ & 1 & 78.1\\
SiBT2 & p$^+$ & 2 & 38.0\\
SiBT2 & n$^+$ & 1 & 44.8\\
SiBT2 & n$^+$ & 2 & 31.8\\
\hline
SiBT3 & p$^+$ & 1 & 59.0\\
SiBT3 & p$^+$ & 2 & 50.8\\
SiBT3 & n$^+$ & 1 & 44.6\\
SiBT3 & n$^+$ & 2 & 41.3\\
\hline
\end{tabular}
\end{table}

Despite this reduction, the SNR remained above 30 for all operational ASICs.

\begin{figure}[htbp]
\centering
\includegraphics[width=.95\textwidth]{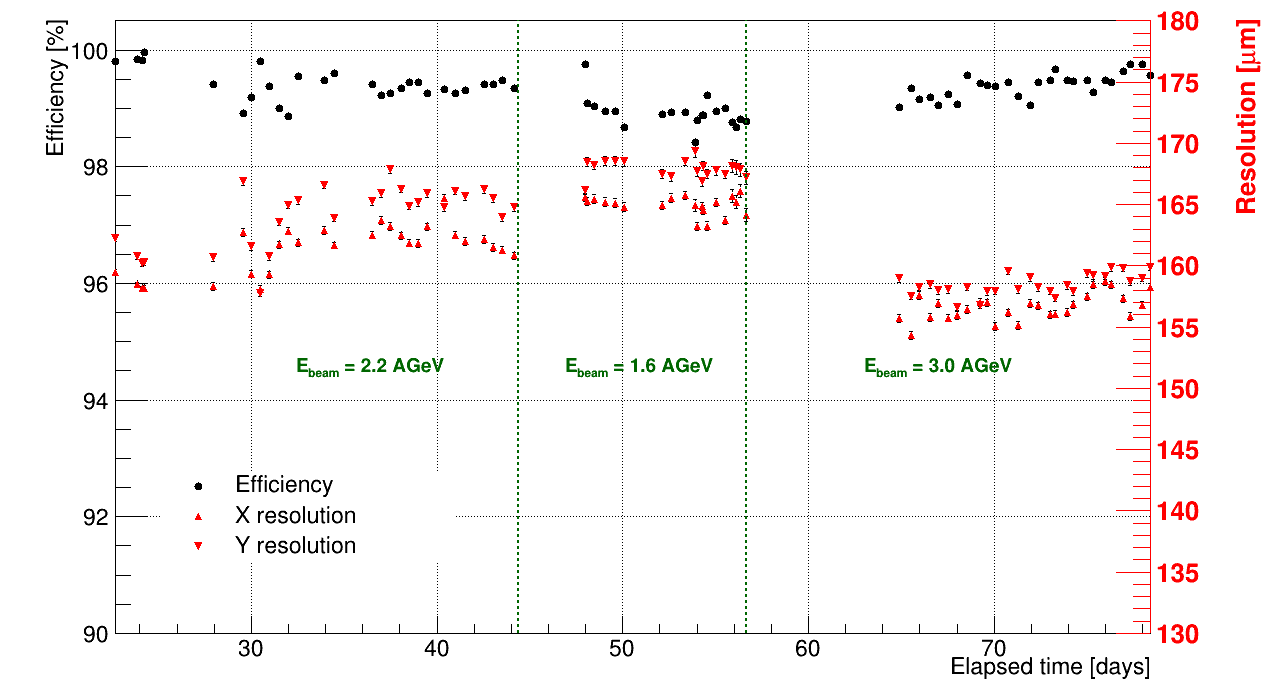}
\caption{SiBT2 tracking efficiency (left axis) and spatial resolution in $X$ and $Y$ (right axis) as a function of the elapsed time since the start of Run 9, for the $2.2$, $1.6$, and $3.0$~AGeV beam-energy periods.\label{fig:eff_res_vs_time}}
\end{figure}

To study whether the accumulated radiation damage affects tracking performance, the procedure of Section~\ref{sec:eff_res} was applied to all 76 runs of Run 9. Figure~\ref{fig:eff_res_vs_time} shows the resulting SiBT2 tracking efficiency (left axis) and spatial resolution (right axis) as a function of elapsed time since the start of the run, for the three periods of data taking at different beam energies.

\noindent Table~\ref{tab:eff_res_summary} summarizes the mean values for each period.

\begin{table}[htbp]
\centering
\caption{Mean SiBT2 tracking efficiency and spatial resolution for the three beam-energy periods of Run 9. Uncertainties are statistical.\label{tab:eff_res_summary}}
\smallskip
\begin{tabular}{c|c|c|c}
\hline
Energy AGeV & Efficiency \% & $\sigma_x$ $\mu$m & $\sigma_y$ $\mu$m \\
\hline
2.2 & $99.33 \pm 0.09$ & $161.26 \pm 0.36$ & $163.87 \pm 0.49$ \\
1.6 & $98.92 \pm 0.06$ & $164.92 \pm 0.17$ & $167.88 \pm 0.16$ \\
3.0 & $99.38 \pm 0.04$ & $156.62 \pm 0.21$ & $158.45 \pm 0.17$ \\
\hline
\end{tabular}
\end{table}
\textbf{Discussion.}
No monotonic decrease of the measured SiBT2 tracking efficiency is observed over the course of the run, despite the radiation damage accumulated by the SiBT sensors. This is illustrated by comparing the first and last periods: the mean efficiency in the final $3.0$~AGeV period, $(99.38\pm0.04)\%$, is not lower than in the initial $2.2$~AGeV period, $(99.33\pm0.09)\%$, although the accumulated damage was largest by then (area-averaged $\Phi_{eq}\approx1.3\times10^{12}~\mathrm{n_{eq}/cm^2}$, locally up to $\approx2.6\times10^{13}~\mathrm{n_{eq}/cm^2}$ in the beam core).

Within each period, a linear fit of efficiency versus elapsed time gives slopes of $(-0.005\pm0.014)$, $(-0.042\pm0.018)$, and $(+0.037\pm0.007)\,\%/\text{day}$ for the $2.2$, $1.6$, and $3.0$~AGeV periods, respectively; these residual trends have different signs and may reflect variations in beam tuning, trigger logic, and front-end-electronics (FEE) conditions following changes in beam energy, rather than a monotonic effect of accumulated fluence. We note that beam energy and elapsed exposure time are correlated, so a small fluence-dependent effect cannot be entirely excluded. However, its magnitude is clearly smaller than the beam-energy dependence discussed next.

The spatial resolution shows a clear dependence on beam energy, while the tracking efficiency varies only slightly between the three periods (Table~\ref{tab:eff_res_summary}): the resolution improves from $\sigma_x=164.9\,\mu\text{m}$, $\sigma_y=167.9\,\mu\text{m}$ at $1.6$~AGeV to $\sigma_x=156.6\,\mu\text{m}$, $\sigma_y=158.5\,\mu\text{m}$ at $3.0$~AGeV, while the efficiency changes from $(98.92\pm0.06)\%$ to $(99.38\pm0.04)\%$. Since the quoted resolution is a convolution of the intrinsic SiBT2 spatial resolution with the pointing uncertainty of the two-point SiBT1--SiBT3 track, this trend is naturally explained by reduced multiple Coulomb scattering of the beam ions, over the $\sim2$~m SiBT1--SiBT3 lever arm, at higher momentum. The fringe magnetic field of the BM@N analyzing magnet in this region may further contribute to the energy dependence, since the bias introduced by the straight-line approximation discussed in Section~\ref{sec:eff_res} is also expected to depend on momentum.

\textbf{Limitations.}
The efficiency and spatial resolution were evaluated for SiBT2 only, using SiBT1 and SiBT3 as reference stations. Inefficiency of the reference stations reduces the number of selected tracks but does not directly affect the measured SiBT2 efficiency. Thus, the present method does not provide an independent efficiency measurement for the three-station system.

The track-unbiased selection requires exactly one hit on each reference station and a reconstructed vertex within $\pm2\sigma$ of the most probable position.  The fraction of retained events ranges from $\approx53\%$ to $\approx75\%$ across the analysed runs, with no monotonic decrease over Run~9. At higher beam intensities, increased hit multiplicity on the reference stations is expected to further reduce this fraction and make track-to-hit association more ambiguous. Therefore, the present results should not be extrapolated to higher-intensity runs without a revised analysis.

The stability of this result against the choice of the matching criterion was verified on three representative runs from the beginning, middle and end of Run~9: tightening the criterion from $dR\le5$ to $dR\le3$ lowers the efficiency by $0.6$--$1.1\%$ and loosening it to $dR\le10$ raises it by $0.2$--$0.6\%$; the lowest value is consistently obtained in the middle, $1.6$~AGeV period, and the efficiency stays above $97.8\%$.

\section{Conclusions}
The Silicon Beam Tracker subsystem operated throughout the 2026 $^{124}$Xe campaign at beam energies of~$1.6$,~$2.2$, and~$3.0$~AGeV. The measured SiBT2 tracking efficiency remained above $98.9\%$ in all three periods, with a spatial resolution of approximately $157$--$168~\mu\mathrm{m}$.

Radiation-induced changes were observed in the detectors during the run. The total leakage current of the SiBT3 detector increased by $28.95~\mu\mathrm{A}$, while the charge collection efficiency in the beam center decreased to approximately $68\%$ at the nominal operating voltage of 30 V. Post-run measurements showed that the CCE in the beam center saturates at $\approx93\%$ above $\approx70~\text{V}$. The mean beam-spot SNR over the operational ASICs decreased from approximately 144 to 45, but remained above 30 for all of them.

Within the sensitivity and selection criteria of the applied tracking analysis, no monotonic reduction in the measured SiBT2 tracking efficiency was observed during Run~9. The observed dependence of the spatial resolution on beam energy is consistent with the contribution of multiple Coulomb scattering and with the potential momentum dependence of the bias introduced by the straight-line track approximation in the fringe magnetic field.

The post-run strip leakage-current measurements were used to reconstruct the spatial distribution of accumulated displacement damage in SiBT3. The resulting area-averaged fluence was approximately $1.3\times10^{12}~\mathrm{n_{eq}/cm^2}$, with the estimates from the total current and strip map agreeing within $\sim10\%$. While the map shape is independent of the damage-factor uncertainty, its absolute scale depends on the adopted value of this factor. The values are 1~MeV neutron-equivalent fluences and should not be interpreted as the number of Xe ions per unit area; the absolute scale also depends on the NIEL scaling, which is unverified for Xe ions at these energies, and no independent ion count was available.

This study documents the operational experience of a single heavy-ion campaign rather than a controlled irradiation study. Its quantitative results should therefore be interpreted in the context of the specific detector design and irradiation conditions considered here, rather than used directly to predict the lifetime of sensors with different properties or under different irradiation conditions. Nevertheless, the results demonstrate that thin double-sided silicon strip detectors can retain tracking performance during direct operation in a relativistic heavy-ion beam despite substantial radiation damage and CCE degradation. Increasing the bias voltage may help recover charge collection after irradiation, although this approach remains to be verified under beam conditions. The modular detector design also allows irradiated sensors to be replaced when necessary.
\acknowledgments
We thank the BM@N Collaboration for its support of this work. We are grateful to the Accelerator Division of the Veksler and Baldin Laboratory of High Energy Physics (VBLHEP), JINR, for providing the $^{124}$Xe ion beams used in this study. The authors thank I.A.~Rufanov (VBLHEP, JINR), for providing his developments in SiBT tracking, and D.V.~Dementiev (VBLHEP, JINR), for valuable advice and for his corrections to the text of the manuscript.

% Bibliography

%% [A] Recommended: using JHEP.bst file
\bibliographystyle{JHEP}
\bibliography{biblio.bib}

@article{kapishin2019studies,
title = {Studies of baryonic matter at the {BM@N} experiment ({JINR})},
journal = {Nuclear Physics A},
volume = {982},
pages = {967-970},
year = {2019},
note = {The 27th International Conference on Ultrarelativistic Nucleus-Nucleus Collisions: Quark Matter 2018},
issn = {0375-9474},
doi = {https://doi.org/10.1016/j.nuclphysa.2018.07.014},
url = {https://www.sciencedirect.com/science/article/pii/S0375947418301337},
author = {Mikhail Kapishin}
}

@inproceedings{proceeding,
  author = "Sedykh, Sergey  and  Chemezov, Danil  and  Kapishin, Mikhail and Khabarov, Sergey and Kopylov, Yurii and Martovitsky, Evgenii and Novozhilov, Sergey and Piyadin, Semen Mikhailovich and Streletskaya, Ekaterina and Tarasov, Oleg and Zamyatin, Nikolay",
  title = "{Beam tracker system for the BM@N/NICA experiment}",
  doi = "10.22323/1.468.0124",
  booktitle = "Proceedings of Technology \& Instrumentation in Particle Physics {\textemdash} PoS(TIPP2023)",
  year = 2025,
  volume = "468",
  pages = "124"
}

@article{sibt_siprof_development,
       author = {{Topko}, Yu. A. and {Khabarov}, S.~V. and {Zamyatin}, N.~I. and {Topko}, B.~L. and {Tarasov}, O.~G. and {Zubarev}, E.~V. and {Kopylov}, Yu. A. and {Streletskaya}, E.~A.},
        title = "{The Development of Silicon Beam Tracker and Beam Profilometer at the {BM@N} Experiment}",
      journal = {Physics of Particles and Nuclei},
         year = 2022,
        month = apr,
       volume = {53},
       number = {2},
        pages = {398-402},
          doi = {10.1134/S1063779622020812},
       adsurl = {https://ui.adsabs.harvard.edu/abs/2022PPN....53..398T}
}

@article{bmn_spectrometer,
title = {The {BM@N} spectrometer at the {NICA} accelerator complex},
journal = {Nuclear Instruments and Methods in Physics Research Section A: Accelerators, Spectrometers, Detectors and Associated Equipment},
volume = {1065},
pages = {169532},
year = {2024},
issn = {0168-9002},
doi = {https://doi.org/10.1016/j.nima.2024.169532},
url = {https://www.sciencedirect.com/science/article/pii/S0168900224004583},
author = {S. Afanasiev and G. Agakishiev and E. Aleksandrov and I. Aleksandrov and P. Alekseev and K. Alishina and V. Astakhov and E. Atkin and T. Aushev and V. Azorskiy and V. Babkin and N. Balashov and R. Barak and A. Baranov and D. Baranov and N. Baranova and N. Barbashina and M. Baznat and S. Bazylev and M. Belov and D. Blau and V. Bocharnikov and G. Bogdanova and A. Bolozdynya and E. Bondar and E. Boos and M. Buryakov and S. Buzin and A. Chebotov and D. Chemezov and J.H. Chen and A. Demanov and D. Dementev and A. Dmitriev and J. Drnoyan and D. Dryablov and A. Dryuk and B. Dubinchik and P. Dulov and A. Egorov and D. Egorov and V. Elsha and A. Fediunin and A. Fedosimova and I. Filippov and I. Filozova and D. Finogeev and I. Gabdrakhmanov and A. Galavanov and O. Gavrischuk and K. Gertsenberger and O. Golosov and V. Golovatyuk and P. Grigoriev and M. Golubeva and F. Guber and S. Ibraimova and D. Idrisov and T. Idrissova and A. Iusupova and A. Ivashkin and A. Izvestnyy and V. Kabadzhov and Sh. Kanokova and M. Kapishin and I. Kapitonov and V. Karjavin and D. Karmanov and N. Karpushkin and R. Kattabekov and V. Kekelidze and S. Khabarov and P. Kharlamov and G. Khudaiberdyev and A. Khukhaeva and A. Khvorostukhin and Yu. Kiryushin and P. Klimai and V. Kolesnikov and A. Kolozhvari and Yu. Kopylov and M. Korolev and L. Kovachev and I. Kovalev and I. Kruglova and Yu. Kovalev and I. Kozlov and V. Kozlov and S. Kuklin and E. Kulish and A. Kurganov and V. Kutergina and A. Kuznetsov and E. Ladygin and D. Lanskoy and N. Lashmanov and I. Lebedev and V. Lenivenko and R. Lednicky and V. Leontiev and D. Lyapin and E. Litvinenko and Y.G. Ma and A. Makankin and A. Makhnev and A. Malakhov and M. Mamaev and A. Martemianov and E. Martovitsky and K. Mashitsin and M. Merkin and S. Merts and S. Morozov and Yu. Murin and K. Musaev and G. Musulmanbekov and A. Myasnikov and D. Myktybekov and R. Nagdasev and S. Nemnyugin and D. Nikitin and S. Novozhilov and Kh. Olimov and K. Olimov and V. Palichik and P. Parfenov and I. Pelevanyuk and D. Peresunko and S. Piyadin and M. Platonova and V. Plotnikov and D. Podgainy and N. Pukhaeva and F. Ratnikov and S. Reshetova and V. Rogov and I. Romanov and I. Rufanov and P. Rukoyatkin and M. Rumyantsev and T. Rybakov and D. Sakulin and S. Sedykh and D. Serebryakov and A. Shabanov and I. Segal and A. Semak and S. Sergeev and A. Serikkanov and A. Sheremetev and A. Sheremeteva and A. Shchipunov and M. Shitenkov and M. Shopova and V. Shumikhin and A. Shutov and V. Shutov and M. Shodmonov and I. Slepnev and V. Slepnev and I. Slepov and A. Smirnov and T. Smolyanin and A. Solomin and A. Sorin and V. Sosnovtsev and V. Spaskov and A. Stavinskiy and V. Stekhanov and Yu. Stepanenko and E. Streletskaya and O. Streltsova and M. Strikhanov and E. Sukhov and D. Suvarieva and G. Taer and A. Taranenko and N. Tarasov and O. Tarasov and P. Teremkov and A. Terletsky and O. Teryaev and V. Tcholakov and V. Tikhomirov and A. Timoshenko and O. Tojiboev and N. Topilin and T. Tretyakova and V. Troshin and A. Truttse and I. Tserruya and V. Tskhay and I. Tyapkin and V. Ustinov and V. Vasendina and V. Velichkov and V. Volkov and A. Voronin and A. Voronin and N. Voytishin and B. Yuldashev and V. Yurevich and N. Zamiatin and M. Zavertyaev and S. Zhang and I. Zhavoronkova and V. Zhezher and N. Zhigareva and A. Zinchenko and A. Zubankov and E. Zubarev and M. Zuev}
}

@misc{ideas,
  author = {{Integrated Detector Electronics AS (IDEAS)}},
  title = {{IDEAS} Integrated Circuits},
  howpublished = {\url{https://ideas.no/ideas-ic-products/}}
}

@article{shafronovskaia2014method,
    author = "Zamyatin, N. I. and Cheremukhin, A. E. and Shafronovskaya, A. I.",
    title = "{Measuring fluence of fast neutrons with planar silicon detectors}",
    doi = "10.1134/S1547477117050156",
    journal = "Phys. Part. Nucl. Lett.",
    volume = "14",
    number = "5",
    pages = "762--777",
    year = "2017"
}

@phdthesis{moll_phd,
    author = "Moll, Michael",
    title = "{Radiation damage in silicon particle detectors: Microscopic defects and macroscopic properties}",
    reportNumber = "DESY-THESIS-1999-040",
    school = "Hamburg U.",
    year = "1999"
}

@incollection{Ivanova2019,
  author = {Ivanova, Y.A. and Khabarov, S.V. and Kovalev, Y.S. and Zamyatin, N.I.},
  title = {Design of the front-end electronics based on the multichannel {IDEAS} {ASICs} for silicon and {GEM} detectors},
  booktitle = {CEUR Workshop Proceedings},
  publisher = {CEUR-WS},
  year = {2019},
  pages = {212-218},
}

@misc{srniel,
    author = "Boschini, M. J. and Rancoita, P. G. and Tacconi, M.",
    title = "{SR-NIEL-7 Calculator: Screened Relativistic (SR) Treatment for NIEL Dose, Nuclear and Electronic Stopping Power Calculator (version 11.1)}",
    howpublished = {\url{https://www.sr-niel.org/} (accessed 2026-11)},
    year = "2014"
}

\end{document}